\documentclass[3p,times,procedia]{elsarticle}
\usepackage{nupha_ecrc}
\usepackage [autostyle, english = american]{csquotes}
\MakeOuterQuote{"}

\volume{00}

\firstpage{1}

\journalname{Nuclear Physics A}

\runauth{}

\jid{nupha}

\jnltitlelogo{Nuclear Physics A}

\usepackage{amssymb}

\usepackage[figuresright]{rotating}

\newcommand{\RAA}{R_{AA}}
\newcommand{\Rpa}{R_{pA}}
\newcommand{\vn}{v_n}

\begin{document}

\begin{frontmatter}



\dochead{XXVIIth International Conference on Ultrarelativistic Nucleus-Nucleus Collisions\\ (Quark Matter 2018)}

\title{$D^0$-Meson $R_{AA}$ in PbPb Collisions at $\sqrt{s_{NN}}$ = 5.02 TeV and Elliptic Flow in pPb Collisions at $\sqrt{s_{NN}}$ = 8.16 TeV with CMS}


\author{Zhaozhong Shi, on behalf of the CMS Collaboration}

\address{Laboratory for Nuclear Science, Massachusetts Institute of Technology, Cambridge, MA USA 02139}

\begin{abstract}
The study of charm production in heavy-ion collisions is considered an excellent probe for the properties of the hot and dense medium created in heavy-ion collisions. Measurements of $D^0$-meson nuclear modification factor can provide strong constraints into the mechanisms of in-medium energy loss and charm flow in the medium. The measurement of $D^0$-meson elliptic flow in pPb collisions helps us understand the strength of charm quarks coupling to significantly reduced systems which demonstrate hydrodynamic properties. In this paper, the measurements of the $D^0$-meson nuclear modification factor in PbPb collisions at 5.02 TeV together with the new measurement of $D^0$-meson elliptic flow in high multiplicity pPb collisions at 5.02 TeV using the two-particle correlation method will be presented.
\end{abstract}

\begin{keyword}
Heavy Flavor, Nuclear Modification Factor, Energy Loss Mechanism, Elliptic Flow, Small Systems

\end{keyword}

\end{frontmatter}


\section{Introduction}

Heavy flavor quarks, such as charm quarks and bottom quarks, are excellent hard probes to study the internal structure and medium properties of the quark-gluon plasma (QGP). Because of their high mass, which are in the order of a few GeV, heavy quarks are created in hard scattering processes in the early stage of collisions. In addition, they have long thermal relaxation times, large diffusion coefficients, and retain their identities when propagating through the QGP medium~\cite{HFQPhysics}. However, heavy quarks lose a significant fraction of their initial energy when they travel through the QGP medium~\cite{HQEnergyLoss}, like the light quarks. Hence, with heavy quarks probes, one can also study the energy loss mechanisms inside the QGP medium.  

In a simplified schematization, there are two different pictures that describe the internal structure of QGP and the energy loss mechanism of heavy quark in the QGP medium. One, perturbative QCD (pQCD), assumes that the coupling of the constituents of the QGP is weak. Therefore, in the pQCD picture, the QGP is made of weakly coupled quasiparticles. Heavy quarks scatter off the constituents incoherently when propagating through the QGP medium. There are two energy loss mechanisms: collisional energy loss and radiative energy loss \cite{HFQPhysics}. 
The other picture, AdS/CFT, takes the strong coupling limit. In this picture, QGP behave like liquid and heavy quarks scatter off the constituents coherently in the QGP medium. The AdS/CFT model applies holographic drag force \cite{AdSCFTDragForce} to calculate the energy loss of heavy quark \cite{AdSCFTHQ} in the QGP medium.

The most common experimental observables to study the production of a particle are the nuclear modification factor $\RAA$ and $\Rpa$, and the anisotropic flow $\vn$. The nuclear modification factor reflects the energy loss, and can help understand the flavor dependence of parton energy loss \cite{FlavorEloss,DeadCone}. Anisotropic flow sheds light on how much the quarks are coupled to a possibly hydrodynamic medium around them~\cite{DpPbv2Paper}. This document presents the latest CMS results on the nuclear modification factor $\RAA$~\cite{DRAAPaper} and elliptic flow $v_2$ in high multiplicity 8.16 TeV pPb \cite{DpPbv2Paper} collisions, for fully reconstructed prompt $D^0$ mesons. 

\section{Analysis Techniques}
\subsection{Online Event Selections}

Dedicated hardware level 1 (L1) trigger and software trigger (HLT) are used to select online $D^0$ meson events. The L1 trigger uses a jet algorithm with online background subtraction, while at the HLT level several single track and $D^0$-meson selections are deployed. The HLT $D^0$ mesons selections are based on reconstructing $D^0$ mesons with loose selections based on the $D^0$-meson vertex displacement, and are used to enhance the amount of high $p_T$ D mesons stored. In the pPb data, in addition, a high multiplicity requirement is added, to select high multiplicity pPb events with multiplicity comparable to peripheral PbPb collisions.

\subsection{$D^0$-meson Reconstruction}

The $D^0$ meson is reconstructed in the decay channel $D^0 \rightarrow K^- \pi^+$ at mid-rapidity region $|y| < 1$. The branching ratio for this channel is about 3.89\% \cite{PDG}, the fragmentation fraction $f(c\rightarrow D^0) \simeq 58.8 \%$ from ZEUS ($\gamma p$) HERA II results \cite{ZEUS}, and $c \tau \simeq 120$ $\mu m$ \cite{PDG}. To reconstruct $D^0$ mesons offline, without particle identification (PID), the secondary $D^0$ meson vertex is reconstructed with a pair of oppositely charged tracks, and several selections on the decay topology are applied: on the pointing angle $\alpha$ between the sum of two track three momentum vectors and the decay length, on the 3D decay length significance, on the secondary vertex probability, and on the distance of closest approach (DCA) \cite{RAATalk}.

\subsection{Prompt Fraction Determination}

The prompt fraction of $D^0$ mesons is determined in a data-driven way, and based on the fact that the average DCA for the non-prompt $D^0$ mesons is larger than for prompt $D^0$ mesons. The DCA distributions in data are fitted using prompt and non-prompt $D^0$ mesons Monte Carlo templates to extract the prompt fraction \cite{DRAAPaper}. To measure the prompt $D^0$-meson nuclear modification factor and elliptic flow, the $D^0$-meson spectrum is corrected with the prompt fractions calculated separately for each data sample.

\subsection{Yield Extraction}

The raw $D^0$-meson yield is extracted by fitting the invariant mass ($m_{inv}$) distribution of the two tracks with a double Gaussian for the signal component, a third order polynomial for the background component, and a single Gaussian for the $K - \pi$ swapped component (candidates with wrong mass assignment). As an example, the fits for $D^0$ mesons with 5 GeV/c $ < p_T <$ 6 GeV/c in pp and PbPb are shown in Figure~\ref{fig:SignalExtraction}:

\begin{figure}[hbtp]
\begin{center}
\includegraphics[width=0.30\textwidth]{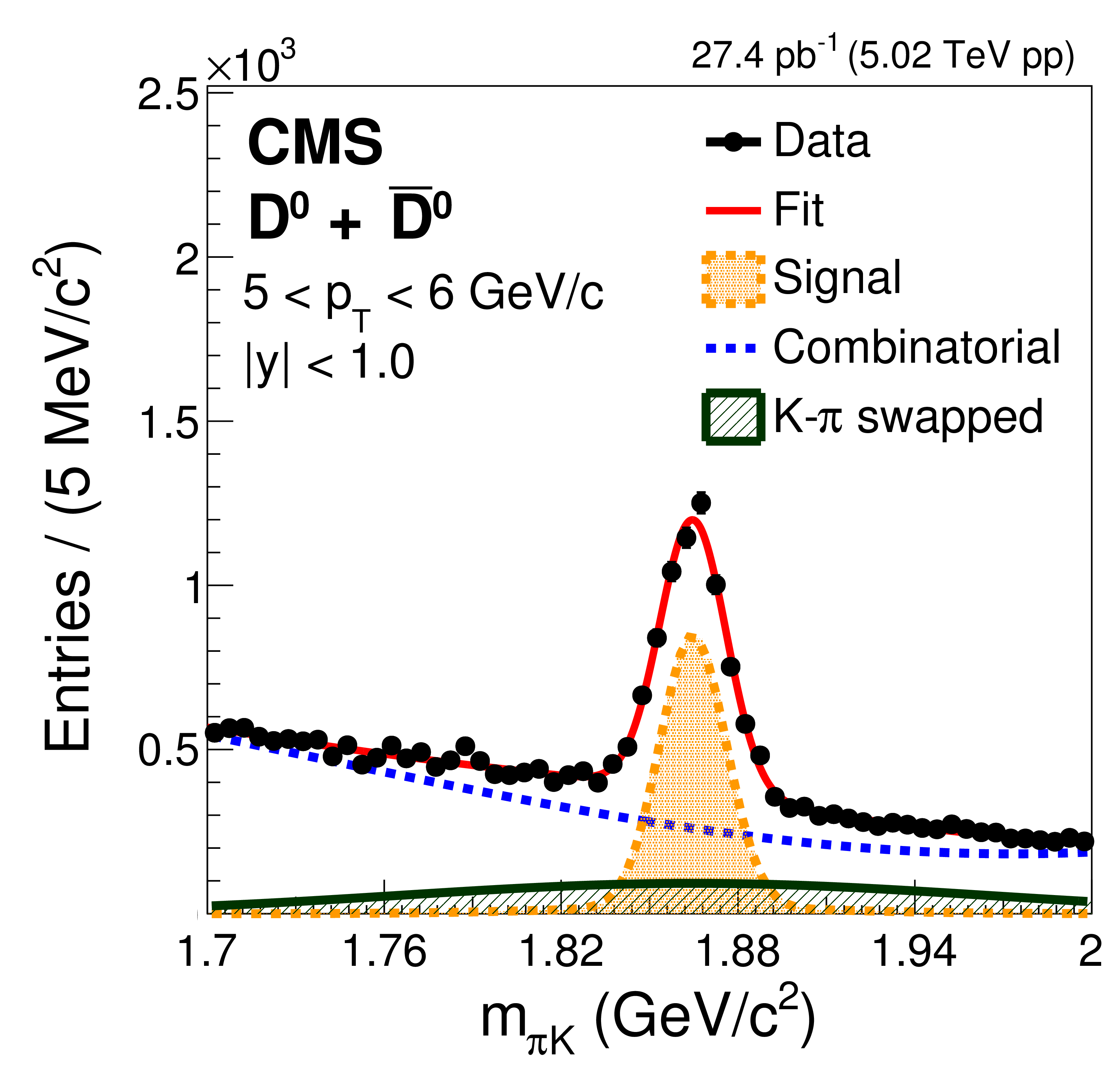}
\includegraphics[width=0.30\textwidth]{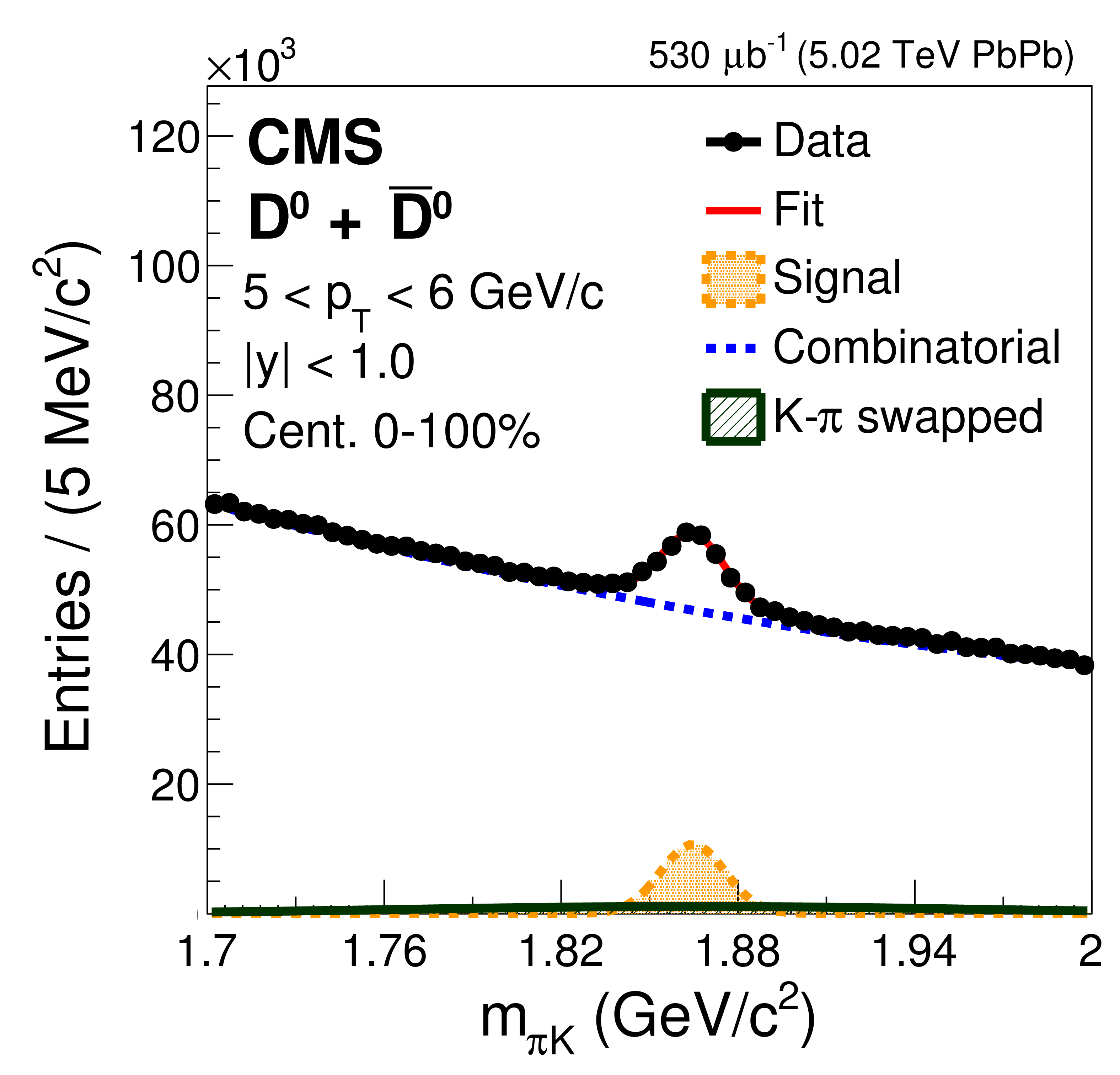}
\caption{Examples of $D^0$ invariant mass distributions in pp (left) and PbPb (right) data \cite{DRAAPaper}.}
\label{fig:SignalExtraction}
\end{center}
\end{figure}

\subsection{$D^0$-Meson Signal Elliptic Flow Determination}

The azimuthal distribution of particles produced in a collision can be described by Fourier series $\frac{dN}{d\phi} \propto 1 + 2 \sum v_n \cos[n(\phi - \psi_n)]$ \cite{FlowPaper}. The second order Fourier harmonics $v_2$ is called the elliptic flow. Here the two-particle correlation method \cite{TPM} is used to extract the elliptic flow. The elliptic flow analysis uses the same techniques as the nuclear-modification-factor analysis to extract the $D^0$-meson yield. First, each $D^0$ meson candidate is paired with all charged tracks produced in the same event, with a pseudorapidity gap $|\Delta\eta| = 1$, to create two-particle correlation distributions. The next step is to perform Fourier fits on the two-particle correlation distributions to extract $V_{2\Delta} (p_T^{D^0}, p_T^{assoc})$. Then, for pPb collision only, we subtract $V_{2\Delta}$ measured in low multiplicity events from that measured in high multiplicity events, to reduce the non-flow contributions. We obtain the elliptic flow $v_2$ from $V_{2\Delta}$: $v_2(p_T) = \frac{V_{2\Delta} (p_T^{D^0}, p_T^{assoc})}{\sqrt{V_{2\Delta} (p_T^{assoc}, p_T^{assoc})}}$. Finally, a simultaneous fit to the $D^0$-meson $m_{inv}$ and $v_2$ vs $m_{inv}$ is performed, to determine the signal component $v_2^S (m_{inv})$ \cite{DpPbv2Paper}.

\section{Results}

\subsection{Prompt $D^0$-Meson $\RAA$ in PbPb Collisions}

After correcting the $D^0$-meson yield by luminosity, efficiencies, and prompt fraction in pp and PbPb, we get the prompt $D^0$-meson nuclear modification factor in $\RAA$ in PbPb collisions and compare it with different particle species and theoretical models shown in Figure~\ref{fig:D0RAA}:

\begin{figure}[hbtp]
\begin{center}
\includegraphics[width=0.30\textwidth]{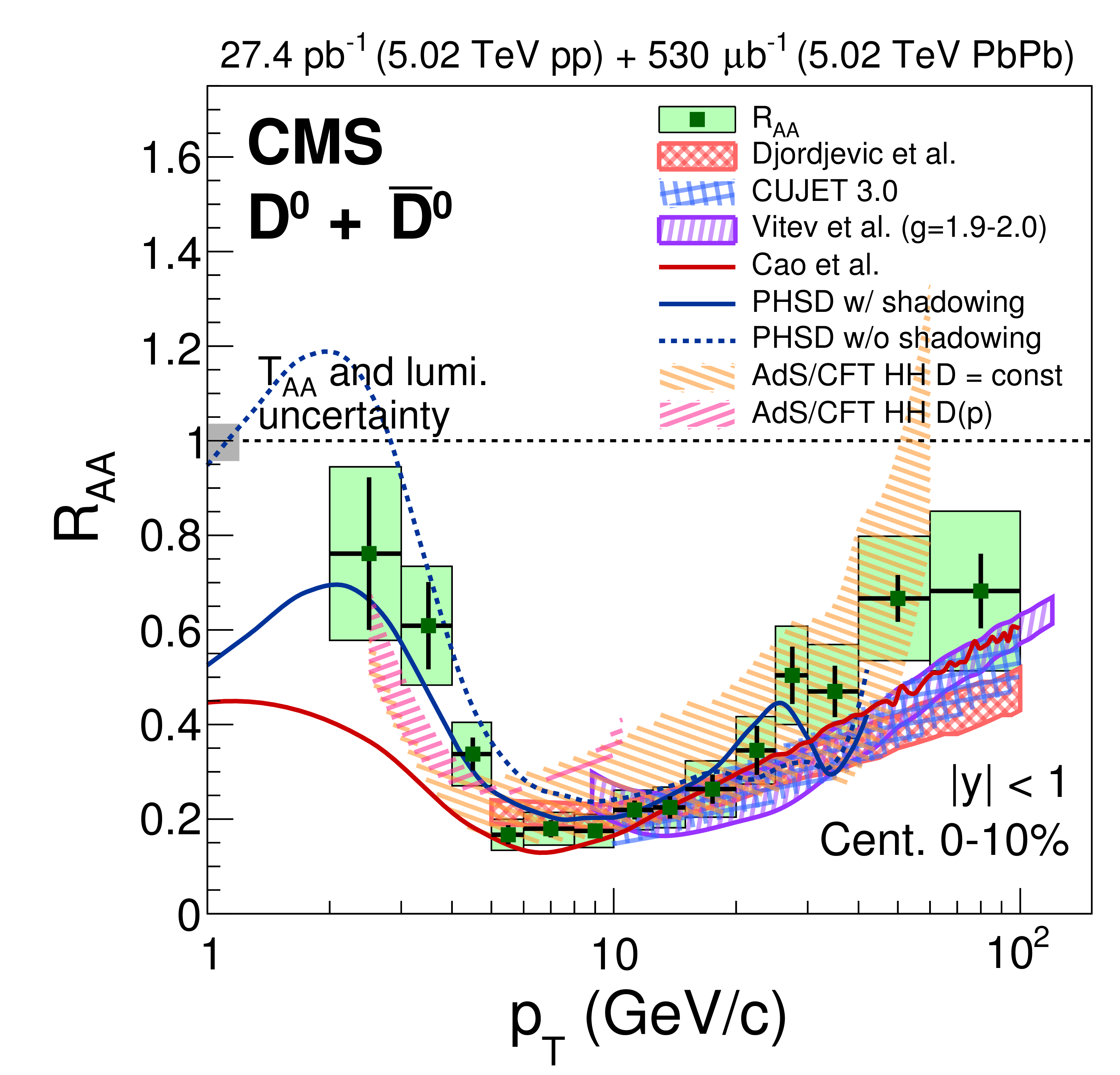}
\includegraphics[width=0.30\textwidth]{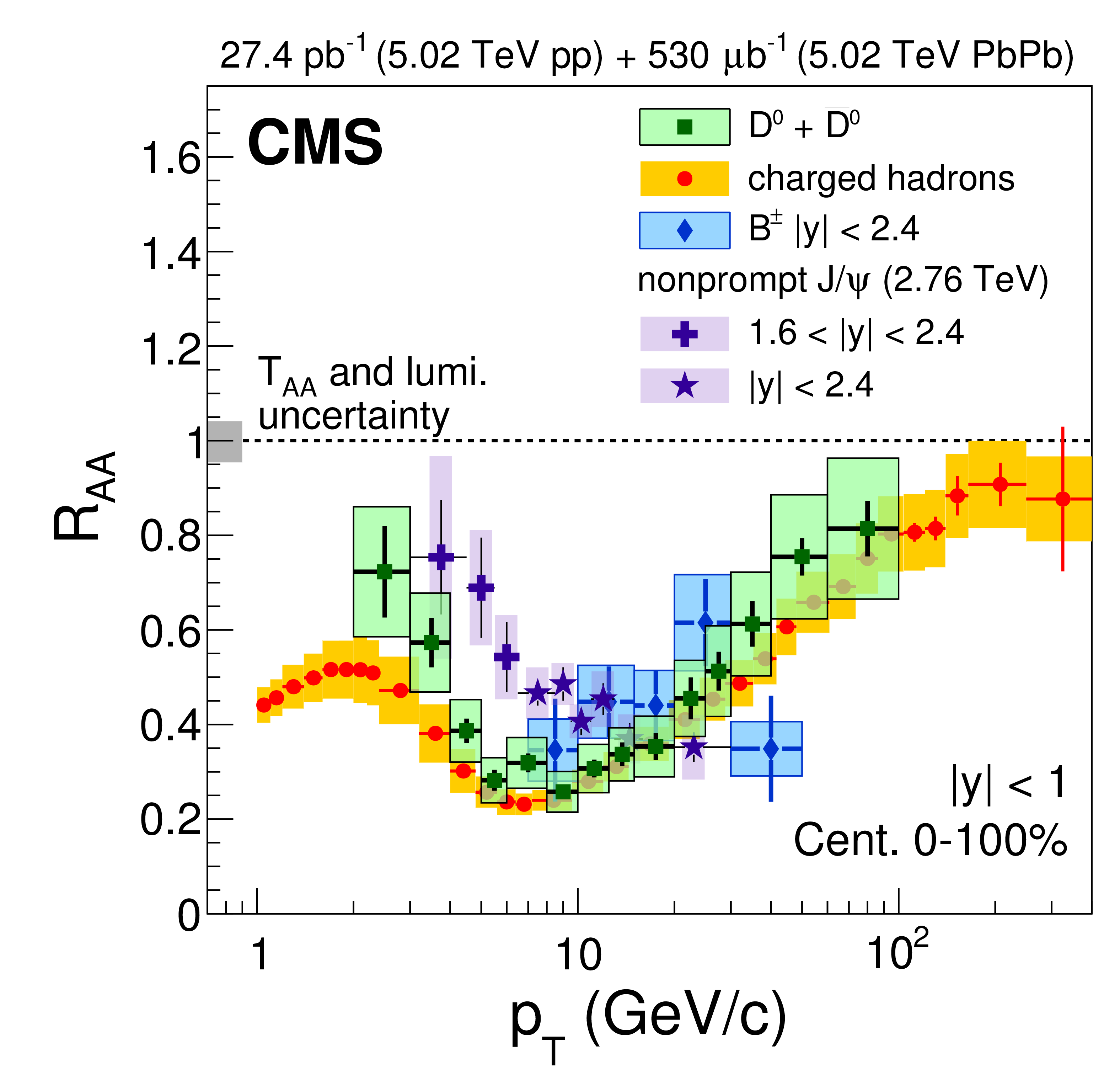}
\caption{The left plot shows the $D^0$-meson $R_{AA}$ vs $p_T$ in PbPb collisions with centrality of 0 -- 10\% and the comparison with various theoretical models~\cite{DRAAPaper}. The right plot shows the $R_{AA}$ of $D^0$ mesons, charged hadrons, $B^{\pm}$ mesons, and non-prompt $J/\psi$ mesons with centrality of 0 -- 100\%~\cite{DRAAPaper}.}
\label{fig:D0RAA}
\end{center}
\end{figure}

The results indicate charm quarks losing a significant fraction of energy in the QGP medium. The $\RAA$ is minimal near $p_T=10$ GeV/c and then increases. At high $p_T$, both pQCD and AdS/CFT predictions are in reasonable agreement with our $\RAA$ results. At low $p_T$, PHSD with shadowing \cite{PHSD} describes our data better. In addition, the suppression of $D^0$ mesons and non-prompt $J/\psi$ mesons is smaller than charged particles. At high $p_T$, $D^0$-meson $\RAA$ is similar to charged particles $\RAA$. The non-prompt $J/\psi$-meson $\RAA$ is higher than the $D^0$-meson $\RAA$ for $p_T \lesssim 15$ GeV/c.

\subsection{Prompt $D^0$-Meson Elliptic Flow in High Multiplicity pPb Collisions}

\begin{figure}[hbtp]
\begin{center}
\includegraphics[width=0.515\textwidth]{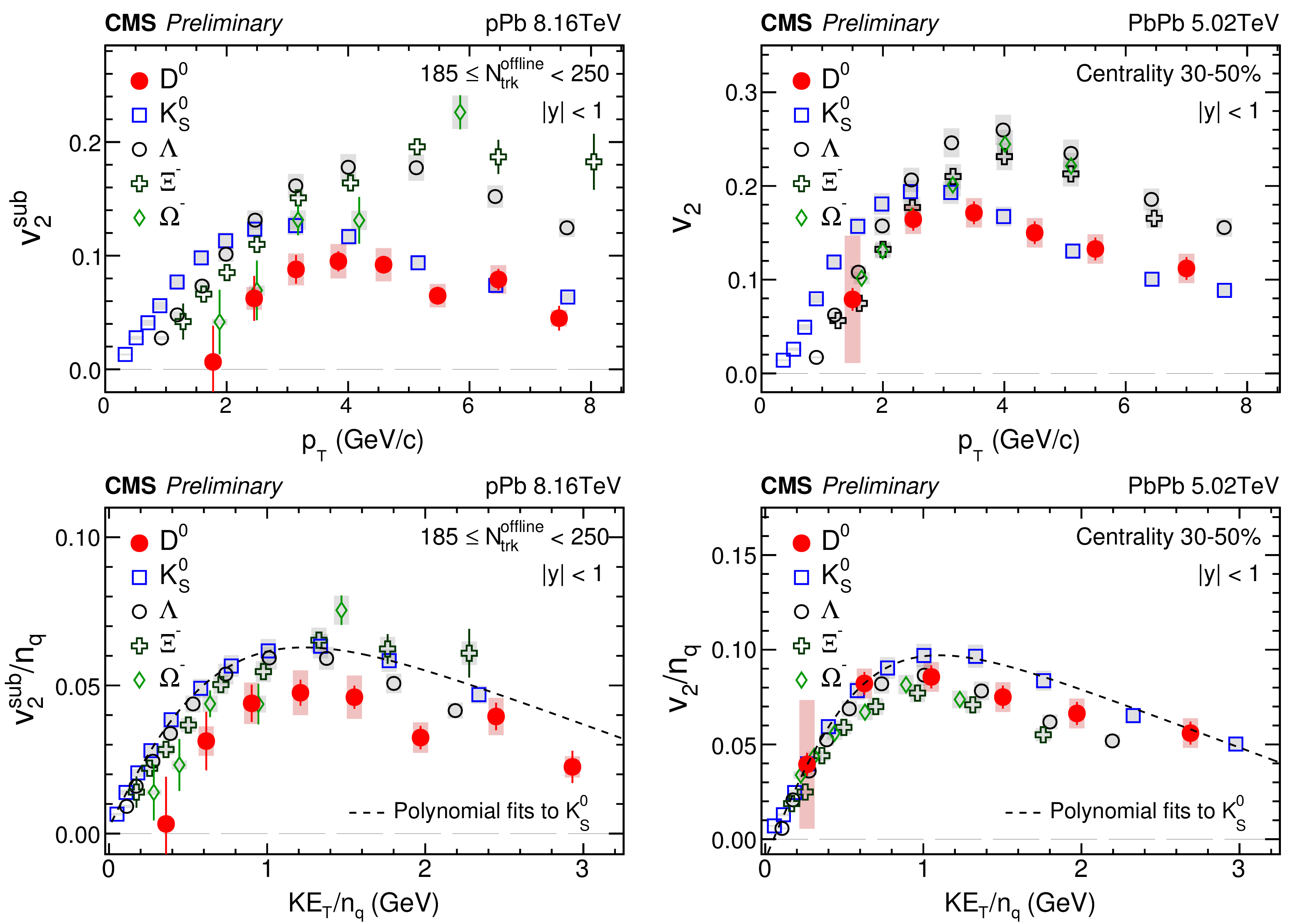}
\caption{The top plots show our measurements of $D^0$, $K_S^0$, $\Lambda$, $\Xi^-$, and $\Omega^-$ $v_2$ vs $p_T$ in pPb (left) and PbPb (Right). The bottom plots, motivated by NCQ scaling \cite{NCQ}, show $v_2/n_q$ vs transverse energy $KE_T/n_q$ in pPb (left) and PbPb (right) \cite{DpPbv2Paper}.}
\label{fig:v2Plots}
\end{center}
\end{figure}

Figure~\ref{fig:v2Plots} shows the prompt $D^0$-meson elliptic flow ($v_2$) result for pPb collisions, compared to the similar results in PbPb collisions. This is the first measurement of $D^0$-meson $v_2$ in pPb collisions.  A significant $v_2$ is observed. The $D^0$ mesons have smaller $v_2$ than strange hadrons in pPb collisions. This suggests that charm quarks do not couple to the small system as strongly as light flavor quarks. In addition, $D^0$-meson $v_2$ in pPb is smaller than in PbPb collision. Finally, in the $v_2/n_q$ vs $KE_T/n_q$ plots, $D^0$ mesons have similar behavior to light flavor hadrons scaled by the number of constituent quarks (NCQ) in PbPb while it is significantly lower than strange hadrons in pPb, which again shows that charm quarks couple to small systems more weakly than light flavor quarks.

\section{Summary}

We have presented the CMS measurements of $D^0$-meson $\RAA$ in PbPb and $v_2$ in pPb collisions. A suppression of $D^0$-meson production is observed from $p_T = $ 2 to 100 GeV/c, with a minimal $R_{AA} \simeq 0.25$ near 10 GeV/c. The first measurement of $D^0$-meson $v_2$ in pPb shows a significant $D^0$-meson $v_2$ in high multiplicity pPb collisions. $D^0$-meson $v_2$ is smaller than that of strange hadrons in pPb collisions, suggesting a weaker coupling of charm quarks to the small system created in pPb collisions compared to strange quarks. 

\section{Acknowledgement}

This work is supported by United States Department of Energy Nuclear Physics Program. We would like to thank the Quark Matter 2018 Conference organizers for giving us an opportunity to present this work. 

\label{}





\bibliographystyle{elsarticle-num}
\bibliography{<your-bib-database>}



\end{document}